# Evidence of natural hybridization and introgression between *Medicago ciliaris* and *Medicago intertexta*

Chérifi Khalil

Laboratory of Biotechnology and Valorization of Natural Resources, Faculty of sciences, Ibn Zohr University, P. O. Box 8106, 80000 Agadir, Morocco

***Abstract***— *The present study, investigated some reproductive and fertility parameters in some wild populations, originating from the North Tunisia (4 populations of Medicago ciliaris and 3 populations of Medicago intertexta). Previous finding revealed that these species are genetically distinct and easily recognized by the number of flowers per inflorescence and pod dimensions. However, biometrical traits and isozyme patterns intermediacy between these two species had detected the existence of a potential spontaneous interspecific hybrid originating from Sedjnane locality in Tunisia. Indeed, the present work has shown significant decrease of pollen fertility and seed production for this population when compared to the others (pollen viability 75%, pollen germinability 8% and pod production=9%). These results suggested a possible natural interspecific hybrid and confirming introgressive hybridization possibility between M. intertexta and M. ciliaris.*

***Keywords***— *Fertility parameters, Natural introgressive hybridization, Genetic diversity, M. ciliaris, M. intertexta, Pollen viability, Pollen fertility.*

## I. INTRODUCTION

For a given species, the reproductive mode is a critically important factor influencing the genetic structure of the natural populations and allows to explain polymorphism. The breeding system affects also many aspects of crop management [1] and contributes to specify the mechanisms of infraspecific differentiation and consequently, the taxonomic value of many forms described within some complex of species [2,3].

Pollen study can be used to differentiate between different species taxonomy and identify possible interspecific hybrids [4,5]. Male sterility derived from irregularity, at any stage, since microsporogenesis to pollen maturation [6]. Pollen abortion of male sterility is an essential requirement in hybrid breeding. This system is verified by several features: Pollen can be characterized on the basis of nuclei number or on the basis of shape and staining behavior [7].The low pollen fertility of the hybrids may be due to chromosome disturbance caused by incompatibility barriers [8].

The genus *Medicago* contains perennial species essentially tetraploid cross-pollinated and annual species largely diploid with a predominant autogamous reproduction mode [9].These annual species are possibly useful to farmers for the production of organic nitrogen and prevention of soil erosion, and also pasture through periods of low production from traditional pasture species [10].

In previous studies, genetic diversity of some natural populations of *Medicago ciliaris* and *Medicago intertexta* (2n=16) originating from Tunisia has been analyzed based on biometrical and isozyme characters. The results have revealed that these species are genetically distinct and easily recognized by the number of flowers per inflorescence and pod dimensions 11]. However, morphological traits and isozyme patterns intermediacy between these two species has been demonstrated in one population originating from Sedjnane region [11]. This population was regarded as being a potential natural interspecific hybrid between the two species.

Characterization of interspecific hybridization in plants has been revealed in various studies [10,12,13,14]. Several methods have been used to reveal hybridity, including intermediate morphology [15,16,17], in situ cross hybridization [18], isozyme [19,20] and DNA [21,22]. The detection of decreased pollen viability or pollen fertility in plant is usually used besides molecular markers for the identification of spontaneous interspecific hybrids [23,24,25]. Pollen viability study is also commonly used in plant breeding for the measure of the reproductive ability of hybrids [24,26].

The aims of this work were to confirm the potential natural hybrid collected from Sedjnane locality. Therefore, fertility comparison study was carried out on several wild populations of *Medicago ciliaris* and *Medicago intertexta* based on pollen viability and germinability and also other aspects of sexual reproduction fitness (pod production, number of seeds per pod and percentage of ovules fertilized and transformed in seeds.

## II. MATERIEL AND METHODS

### 2.1 Plant material

Seven wild populations prospected and collected in native pasture as pods from different natural areas in the North of Tunisia were studied: Four populations of *Medicago ciliaris* and three populations of *Medicago intertexta*. The four ecotypes of *Medicago ciliaris* are originating from the locality of Bizerte (Biz), Jendouba (Jen), Menzel-Bourguiba (Meb) and Korba (Kor). Those of *Medicago intertexta* are collected near Sedjnane (Sed), Mateur (Mat) and Aïn Draham (Aïd).

Seeds collected from randomly selected plant were placed in plastic Petri dishes (90 mm diameter) on filter paper wetted with distilled water. Petri dishes were placed in a precision incubator and maintained in the dark at $20 \pm 0.5°C$. After five days of sowing, one seedling, randomly chosen from each maternal plant, was planted in pot with soil and grown in a greenhouse under uniform conditions.

### 2.2 Fertility analysis

For these experiments, the following parameters related of sexual reproduction fitness and reported by different authors [27,28], were recorded on individual plant:

- Pod production (P/F) where P = number of pod and F = number of flower.

- Average number of seeds per pod S/P where S = number of seed and P = number of pod

- Index of seed maturation which is the proportion of ovules fertilized and transformed in seeds S/O where S/O is equal to ratio of S/P to o/O (o/O is the mean number of ovule (o) per ovary (O). This parameter gives an overview of the fertilization effectiveness [29].

### 2.3 Pollen viability and germinability

Pollen viability and germinability analysis were commonly used in plant breeding for various species because of their facility, rapidity and low financial cost and the dependability of the technique [30].

Pollen were collected from the anthers of flowers in early development stage. The anther was crushed on a slide glass containing a drop of the dye of Alexander stain [31]. When observed under microscope, the pollen grain was considered viable if it turned red or pink while empty pollen had green color. Four counts were made on 4 different fields in each of three microslides of all populations. At least 100 pollens were counted per observation. The percentage of pollen viability was determined as the ratio of the number of viable grains to the total number of grains.

The percentage of germinated pollen was estimated in vitro using the protocol of the hanging drop assay: Three anthers were squeezed on a slide in a drop of aqueous solution with 30 % of sucrose. Then the slide is turned over on a dish containing distilled water. Germination was occurred under obscurity in atmospheric conditions of the laboratory [32]. Germination of pollen grains was noted at the end of 30 minutes of incubation using a photonic microscope system.

### 2.4 Statistical analysis

All values expressed as a percentage were arcsine square root transformed before performing statistical analysis to normalize the data and to homogenize the variance [33]. The data were submitted to a one ways analysis of variance (fixed model) to determine the importance of population effect, followed by a Duncan test. A difference was considered to be statistically significant when $P < 0.05$.

Canonical discriminant analysis, was used to identify the combination of variables that best separate simultaneously all the populations of the two species. This analysis is a multivariate technique that describes the relationship between the parameters by calculating the linear combinations that are maximally correlated [34]. In the present study, the five variables of fertility were introduced as predictors into the discriminant analysis. All statistical analysis were performed with SAS computer program [35] with PROC ANOVA and PROC CANDISC procedures.

## III. RESULTS AND DISCUSSION

### 3.1 Pollen viability and germinability

Pollen viability rate revealed by Alexander coloration, showed significant differences among populations of the tow species (Table 1).

TABLE 1
VARIANCE ANALYSIS AND DUNCAN TEST OF POLLEN VIABILITY PERCENTAGE FOR
*M. ciliaris* AND *M. intertexta* POPULATIONS

| Species | | *Medicago ciliaris* | | | *Medicago intertexta* | | |
|---|---|---|---|---|---|---|---|
| **Variance analysis** | Source | df | MC | F | df | MC | F |
| | Population | 3 | 0.010 | 0.64** | 2 | 0.18 | 34.00* |
| | Error | 16 | 0.001 | | 12 | 0.046 | |
| **Duncan test** | | -------- -------- ---- | | | --------- ---- | | |
| | | Meb  Jen  Biz  Kor | | | Mat  Aïd  Sed | | |
| | | 99.6  99.3  98.7  97.4 | | | 99.6  98.9  75.0 | | |

*df: Degree of freedom; MC: Mean square; F: ratio of variances; \*: significant test; \*\*: high significant test; Values under the same line are not significantly different (p < 0.05)*

On the average, viable pollen represented about 95% of the total pollen grains (Figure 1), however, only the population of Sed of *M. intertexta*, was been characterized by the lowest percentage of viable pollen (75%) compared to the other populations of the tow species in which values was varied from 97,4 % to 99,3 %. For a proper pollination, it is essential to have a high proportion of engorged pollen grains that have the potential for germinating and fertilizing the ovules within the spikelet. These patterns of low pollen viability (relatively low reproductive fitness) observed at this population seem to characterize the most frequent hybrids in plant species.

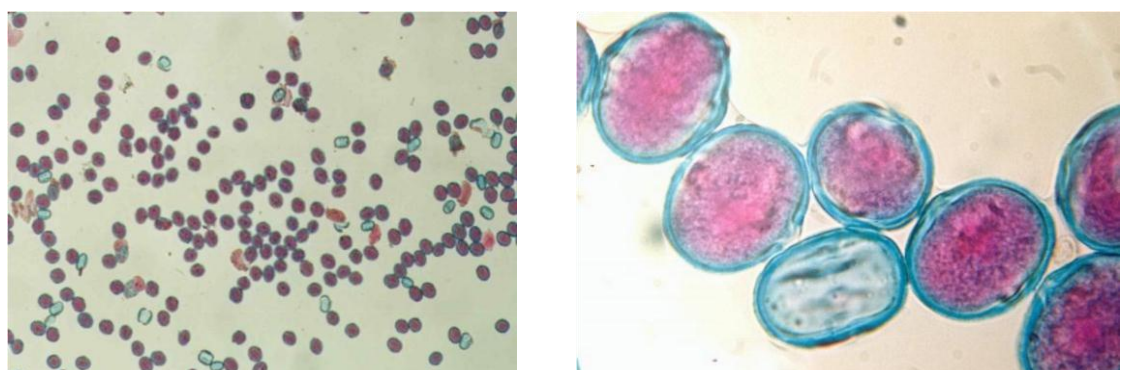

**FIGURE 1 . POLLEN VIABILITY TEST REVEALED BY ALEXANDER COLORATION (X40)**

Furthermore, pollen germinability varied significantly for populations having nearly the same pollen viability (Table 2).

TABLE 2
VARIANCE ANALYSIS AND DUNCAN TEST OF POLLEN GERMINABILITY PERCENTAGE FOR
*M. ciliaris* AND *M. intertexta* POPULATIONS

| Species | | *Medicago ciliaris* | | | *Medicago intertexta* | | |
|---|---|---|---|---|---|---|---|
| **Variance analysis** | Source | df | MC | F | df | MC | F |
| | Population | 3 | 0.070 | 0.62NS | 2 | 0.539 | 16.26** |
| | Error | 16 | 0.120 | | 12 | 0.330 | |
| **Duncan test** | | ---------------------- | | | --------- ---- | | |
| | | Meb  Kor  Jen  Biz | | | Mat  Aïd  Sed | | |
| | | 44.4  45.5  38.5  35.6 | | | 65.2  45.5  08.1 | | |

*df: Degree of freedom; MC: Mean square; F: ratio of variances; NS: non-significant test;*
*\*\*: high significant test; Values under the same line are not significantly different (p < 0.05)*

Aïd population from *M. intertexta* showed de highest percentage of fertility (65,5 %), while Sed, supposed from the same species had the lowest (8,1 %). Although, theoretically the number of pollen grain able to germinate on stigma may be even more lower at this population.

## 3.2 Variability of the fertility parameters

### 3.2.1 *Medicago ciliaris*

For all the studied parameters, the analysis of the variance applied to the data revealed significant or high significant differences between populations (Table 3).

**TABLE 3**
**VARIANCE ANALYSIS AND DUNCAN TEST OF DIFFERENT FERTILITY PARAMETERS P/F, S/P AND S/O CALCULATED FOR POPULATIONS OF *M. ciliaris*.**

| Critères | | P/F° | | | S/P | | | S/O° | | |
|---|---|---|---|---|---|---|---|---|---|---|
| **Variance analysis** | Source | df | MC | F | df | MC | F | df | MC | F |
| | Population | 3 | 0.102 | 10.00** | 3 | 0.653 | 3.24* | 3 | 0.026 | 3.76* |
| | Error | 56 | 0.005 | | 56 | 0.201 | | 56 | 0.007 | |
| **Duncan test** | | --- --------- ---- | | | --------------- --------------- | | | --------------- --------------- | | |
| | | Kor Meb Biz Jen | | | Kor Jen Biz Meb | | | Kor Meb Biz Jen | | |
| | | 0.35 0.26 0.25 0.17 | | | 8.22 7.93 7.92 7.72 | | | 0.95 0.92 0.92 0.91 | | |

*df: Degree of freedom; MC: Mean square; F: ratio of variances; *:significant test;*
*\*\*: high significant test; Values under the same line are not significantly different (p < 0.05)*

The Duncan test (Table 3) indicated that the parameter P/F was the most discriminant. Indeed, tree groups were distinguished. The first was formed by Kor population which showed very high production of pod per flower (P/F= 35%), while Biz and Meb occupied an intermediate position between the most fertile population and Jen, considered as the less productive population with P/F = 17%. Otherwise, the frequency of ovules transformed into seeds (S/O), reflecting the effectiveness of fertilization, varied from 0.91 for the population of Jendouba, to 0.95 for Korba (most fertile). However, there were no significant differences between Meb and Biz, which were close geographically. This finding, observed also in the morphological variability analysis [11], revealed that the important polymorphism observed at this population was correlated to the eco-geographical origin and allowed for the possibility of ecotypes selection well adapted by both their morphology and their reproduction ability.

### 3.2.2 *Medicago intertexta*

High significant differences among populations of *M. intertexta*, were observed for all parameters, except the number of seeds per pod (Table 4).

**TABLE 4**
**VARIANCE ANALYSIS AND DUNCAN TEST OF DIFFERENT FERTILITY PARAMETERS P/F, S/P AND S/O CALCULATED FOR POPULATIONS OF *M. intertexta***

| Critères | | P/F° | | | S/P | | | S/O° | | |
|---|---|---|---|---|---|---|---|---|---|---|
| **Variance analysis** | Source | Df | MC | F | Df | MC | F | Df | MC | F |
| | Population | 2 | 0.087 | 34.61** | 2 | 1.650 | 0.83NS | 2 | 0.624 | 13.08** |
| | Error | 42 | 0.002 | | 42 | 1.981 | | 42 | 0.047 | |
| **Duncan test** | | --------- ---- | | | --------------- | | | --------- ---- | | |
| | | Aïd Mat Sed | | | Aïd Mat Sed | | | Aïd Mat Sed | | |
| | | 0.19 0.18 0.09 | | | 7.26 6.85 6.61 | | | 0.95 0.95 0.68 | | |

*df: Degree of freedom; MC: Mean square; F: ratio of variances; \*\*: high significant test;*
*NS: non-significant test; Values under the same line are not significantly different (p < 0.05)*

The proportion of pod (P/F) observed for this species was generally lower than the one observed for *M. ciliaris*. It ranged from 0.09 to 0.19 for the populations of *M. intertexta* and from 0.17 to 0.35 in the other species. It should be noted that, Sed population was characterized by the lowest fertility (G/F=0.09 and g/O=0.68). The low rate of pollen viability could, at least in part, explained the decline of fertility in this population at self-pollination.

It was found that 92 % of Sed population's pollen did not germinate in vitro at all, and only about 8 % of the pollen germination was normal. By contrast, the germination rate of the remaining populations pollen reached 65,5% under the same conditions.

### 3.2.3    Mixed populations of the two species

The Canonical discriminant analysis, based simultaneously on the global populations of the two species, showed that the first two canonical axes participated in 95.86% of the total discrimination, of which the first axe with 75.54% (Table 5).

**TABLE 5**
**EIGENVALUES FOR THE FIRST THREE CANONICAL VARIABLES RESULTING FROM THE DISCRIMINANT ANALYSIS**

| Canonical variables | Eigenvalue | Percent of variance | Cumulative percent | Canonical correlation |
|---|---|---|---|---|
| Canonical 1 | 35,42 | 75,54 | 75,54 | 0,99 |
| Canonical 2 | 9,53 | 20,32 | 95,86 | 0,95 |
| Canonical 3 | 1,66 | 3,53 | 99,40 | 0,79 |

The first discriminant function is determined by both the viability and fertility of pollen grains and index of seed maturation (S/O) while the second function is mostly defined by the pod production (P/F) (Table 6).

**TABLE 6**
**STANDARDIZED CANONICAL DISCRIMINANT FUNCTION COEFFICIENTS FOR ALL VARIABLES.**

| Canonical variables | Pollen V | Pollen G | P/F | S/O | S/P |
|---|---|---|---|---|---|
| Canonical 1 | 0,33 | **0,95** | -0,60 | **0,97** | **-0,95** |
| Canonical 2 | 0,28 | -0,45 | **1,76** | -0,31 | -0,78 |
| Canonical 3 | 0,48 | **-0,91** | -0,60 | **0,83** | 0,38 |

*V: viability; G: germinability*

The first two discriminant axis clearly separated all the populations into distinct groups (figure 2).

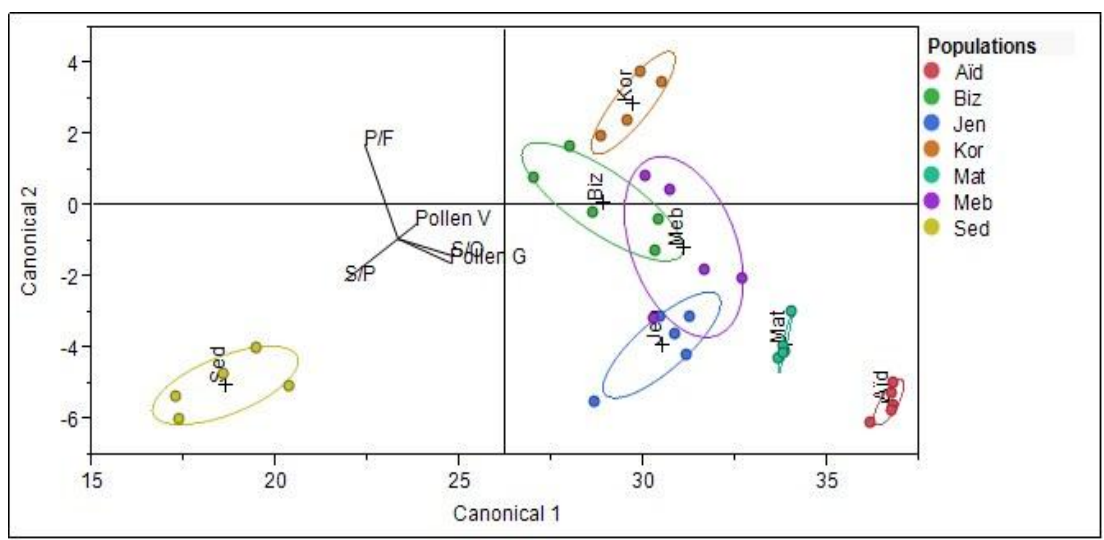

**FIGURE 2. RESULTS OF CANONICAL DISCRIMINANT ANALYSIS OF THE GLOBAL POPULATIONS OF THE TWO SPECIES BASED ON THE FIVE FERTILITY PARAMETERS.**

The population's classification belonging to the species group revealed the same collections in Duncan's multiple range test. Moreover, Sed population was isolated and been characterized by the lowest values for all fertility parameters. This situation often characterized some natural interspecific hybrid [36,37].

By axis 2, *M. ciliaris* ecotypes are well separated regarding to the first canonical axis. This species revealed high variability in pod production between its populations.

The population's divergence depending on the species was not always evident: Jen population of *M. ciliaris* joined those of *M. intertexta* Mat and Aïd along to the second axis.

In accordance with previously published results of the genetic variability analysis of *Medicago ciliaris* and *Medicago intertexta,* fertility comparisons by canonical discriminant analysis confirmed the potential of the natural interspecific hybrid collected from Sedjnane locality. Similar results were observed in several studies for other species [38,39,40,41,13]. These authors reported that spontaneous interspecific hybridizations were also characterized by quantitative traits associated with fertility factors.

The decline of fecundity in the Sed population at self-pollination should be attributed to the irregularity in male gametophyte. This phenotype was very similar to that found in male sterile plants, which can be explained as a condition in which the pollen is not viable and not able to germinate or fertilize normally to produce seeds [42].High meiotic instability might affected pollen grain viability and served as the limiting factor in the process of self-fertilization [43]. Meiotic abnormalities were also reported as responsible for decreased pollen fertility in Alfalfa [44], Lathyrus [44], Bean [45] and Astragalus [45].

Furthermore, the reduced pollen fertility associated with the decreased spike fertility in the male-sterile population may be due to either genetic or environmental factors, such as water deficits or thermal stress. Male sterility commonly results from the interaction between nuclear and cytoplasmic genes, where mutations in mitochondrial DNA were the main source of the sterility [43].

## IV. CONCLUSION

Fertility comparisons of the analyzed populations revealed high variability in reproduction ability. Whereas the hybrid's behaviour of Sedjnane area reacted like an interspecific hybrid between the two species, its fertility was dramatically disturbed. Therefore, individuals of this population observed in the field should be avoided in breeding programs because it may contribute to improve crop breeding and seed propagation effects. Further studies are required, particularly in pollen physiology and cytogenetic investigations associated with observation of fertility problems, to enhance our understanding about the observed natural hybrid. Moreover, development of molecular markers can be used to confirm the genomic contribution of parental taxa to suspected hybrid and its potential implication in the taxonomic problems within the *Medicago* genus.

## REFERENCES

[1] Koelling, J., Coles M. C., Matthews P. D. and Schwekendiek A., 2012, Development of new microsatellite markers (SSRs) for *Humulus lupulus*. Molecular breeding, 30(1), 479-484

[2] Small, E., Warwick S. I. and Brookes B., 1999, Allozyme variation in relation to morphology and taxonomy in *Medicago* sect. Spirocarpos subsect. Intertextae (Fabaceae). Plant systematics and evolution, 214(1-4), 29-47

[3] Kartzinel, R. Y., Spalink D., Waller D. M. and Givnish T. J., 2016, Divergence and isolation of cryptic sympatric taxa within the annual legume *Amphicarpaea bracteata*. Ecology and evolution, 6(10), 3367-3379

[4] Mariani, A., Tavoletti S. and Veronesi F., 1993, Abnormal macrosporogenesis in five alfalfa (*Medicago sativa*) mutants producing 4n pollen. Theoretical and Applied Genetics, 85(6-7), 873-881

[5] Joshi, B. K., Subedi L. P., Gurung S. B. and Sharma R. C., 2014, Pollen and spikelet analysis in f1 rice hybrids and their parents. Nepal Agriculture Research Journal, 8, 120-126

[6] Subbarayudu, S., Naik B. S., Devi H. S., Bhau B. and Khan P. S. S. V., 2014, Microsporogenesis and pollen formation in *Zingiber officinale* Roscoe. Plant systematics and evolution, 300(4), 619-632

[7] Raj, K. G. and Virmani S., 1988, Genetic of fertility restoration of 'WA'type cytoplasmic male sterility in rice. Crop science, 28(5), 787-792

[8] Martinez-Reyna, J. and Vogel K., 2002, Incompatibility systems in switchgrass. Crop Science, 42(6), 1800-1805

[9] Small, E. and Jomphe M., 1989, A synopsis of the genus Medicago (Leguminosae). Canadian Journal of Botany, 67(11), 3260-3294

[10] Brummer, E., Bouton J. and Kochert G., 1995, Analysis of annual *Medicago* species using RAPD markers. Genome, 38(2), 362-367

[11] Cherifi, K., Boussaïd M. and Marrakchi M., 1993, Diversité génétique de quelques populations naturelles de *Medicago ciliaris* (L) Krock et de *Medicago intertexta* (L) Mill. I. Analyse de la variabilité morphologique. Agronomie, 13(10), 895-906

[12] Al Mazyad, P. R. and Ammann K., 1999, Biogeographical assay and natural gene flow. Methods for Risk Assessment of Transgenic Plants, Springer: 95-98

[13] Benabdelmouna, A., Guéritaine G., Abirached-Darmency M. and Darmency H., 2003, Genome discrimination in progeny of interspecific hybrids between Brassica napus and *Raphanus raphanistrum*. Genome, 46(3), 469-472

[14] Rao, S. A., Schiller J., Bounphanousay C., Inthapanya P. and Jackson M., 2006, The colored pericarp (black) rice of Laos. Rice in Laos. International Rice Research Institute, Manila, 175-186

[15] Saltonstall, K., Castillo H. E. and Blossey B., 2014, Confirmed field hybridization of native and introduced *Phragmites australis* (Poaceae) in North America. American journal of botany, 101(1), 211-215

[16] Shu, Z., Zhang X., Yu D., Xue S. and Wang H., 2016, Natural Hybridization between Persian Walnut and Chinese Walnut Revealed by Simple Sequence Repeat Markers. Journal of the American Society for Horticultural Science, 141(2), 146-150

[17] Welker, C. A., Souza-Chies T. T., Longhi-Wagner H. M., Peichoto M. C., McKain M. R. and Kellogg E. A., 2016, Multilocus phylogeny and phylogenomics of *Eriochrysis* P. Beauv.(Poaceae–Andropogoneae): Taxonomic implications and evidence of interspecific hybridization. Molecular phylogenetics and evolution, 99, 155-167

[18] Melo, C., Silva G. and Souza M., 2015, Establishment of the genomic in situ hybridization (GISH) technique for analysis in interspecific hybrids of *Passiflora*. Genet Mol Res, 14(1), 2176-2188

[19] Wagner, I., Maurer W., Lemmen P., Schmitt H., Wagner M., Binder M. and Patzak P., 2014, Hybridization and genetic diversity in wild apple (*Malus sylvestris* (L.) Mill) from various regions in Germany and from Luxembourg. Silvae Genet, 63, 81-94

[20] Yang, Y., Chi D., Cao W., Zeng J., Xue A., Han F. and Fedak G., 2015, Monitoring the introgression of E genome chromosomes into triticale using multicolor GISH. Caryologia, 68(4), 317-322

[21] Sutkowska, A., Boroń P. and Mitka J., 2013, Natural hybrid zone of Aconitum species in the Western Carpathians: Linnaean taxonomy and ISSR fingerprinting. Acta Biologica Cracoviensia Series Botanica, 55(1), 114-126

[22] One, K. T., Tanya P., Muakrong N., Laosatit K. and Srinives P., 2014, Phenotypic and genotypic variability of F 2 plants derived from *Jatropha curcas× integerrima* hybrid. biomass and bioenergy, 67, 137-144

[23] Lihová, J., Kučera J., Perný M. and Marhold K., 2007, Hybridization between two polyploid Cardamine (Brassicaceae) species in north-western Spain: discordance between morphological and genetic variation patterns. Annals of Botany, 99(6), 1083-1096

[24] Bureš, P., Šmarda P., Rotreklova O., Oberreiter M., Burešová M., Konečný J., Knoll A., Fajmon K. and Šmerda J., 2010, Pollen viability and natural hybridization of Central European species of *Cirsium*. Preslia, 82(4), 391-422

[25] Séguin‑Swartz, G., Nettleton J. A., Sauder C., Warwick S. I. and Gugel R. K., 2013, Hybridization between *Camelina sativa* (L.) Crantz (false flax) and North American *Camelina* species. Plant Breeding, 132(4), 390-396

[26] de Paula, C. M. P., Techio V. H., Benites F. R. G. and de Souza Sobrinho F., 2014, Intra-inflorescence pollen viability in accessions of *Brachiaria ruziziensis*. Acta Scientiarum. Biological Sciences, 36(2), 209-213

[27] Dattée, Y., 1976, Analyse quantitative de l'auto et l'interfertilite chez quelques familles la luzerne diploïde. Ann Amelior Plantes, 26(3), 419-441

[28] Chen, W., 2009, Pollination, fertilization and floral traits co-segregating with autofertility in faba bean. Journal of New Seeds, 10(1), 14-30

[29] Stodliard, F., 1986, Pollination, Fertilization and Seed Development in Inbred Lines and F] Hybrids of Spring Faba Beans (*Vicia faba* L.). Plant breeding, 97, 210-221

[30] Cardoso, R. D., Grando M. F., Basso S., Segeren M., Augustin L. and Suzin M., 2009, Caracterização citogenética, viabilidade de pólen e hibridação artificial em gérbera. Horticultura Brasileira, 27, 040-044

[31] Alexander, M., 1969, Differential staining of aborted and non aborted pollen. Stain technology, 44(3), 117-122

[32] Hamadi, H., 2011, Study of Some Characteristics Related to the Sexual Reproduction in Autotetraploid *Vicia narbonensis*. Journal of Agricultural Science, 3(1), 153

[33] Turkington, R., John E., Watson S. and Seccombe‑Hett P., 2002, The effects of fertilization and herbivory on the herbaceous vegetation of the boreal forest in north‑western Canada: a 10‑year study. Journal of Ecology, 90(2), 325-337

[34] Cruz-Castillo, J., Ganeshanandam S., MacKay B., Lawes G., Lawoko C. and Woolley D., 1994, Applications of canonical discriminant analysis in horticultural research. HortScience, 29(10), 1115-1119

[35] SAS, 1990, Statistical Aanalysis System, SAS/STAT user's guide: Version 6, Sas Inst

[36] Hinton, W. F., 1975, Natural hybridization and extinction of a population of *Physalis virginiana* (Solanaceae). American Journal of Botany, 198-202

[37] Rieger, M., Potter T., Preston C. and Powles S., 2001, Hybridisation between *Brassica napus* L. and *Raphanus raphanistrum* L. under agronomic field conditions. Theoretical and Applied Genetics, 103(4), 555-560

[38] Pliszko, A. and Zalewska-Gałosz J., Molecular evidence for hybridization between invasive *Solidago canadensis*. Biological Invasions, 1-6

[39] Gupta, V. and Gudu S., 1991, Interspecific hybrids and possible phylogenetic relations in grain amaranths. Euphytica, 52(1), 33-38

[40] Amand, P. S., Skinner D. Z. and Peaden R. N., 2000, Risk of alfalfa transgene dissemination and scale-dependent effects. Theoretical and Applied Genetics, 101(1-2), 107-114

[41] Guadagnuolo, R., Savova-Bianchi D. and Felber F., 2001, Gene flow from wheat (Triticum aestivum L.) to jointed goatgrass (*Aegilops cylindrica* Host.), as revealed by RAPD and microsatellite markers. Theoretical and applied Genetics, 103(1), 1-8

[42] Stockmeyer, K. and Kempken F., 2006, Biotechnology: Engineered male sterility in plant hybrid breeding. Progress in Botany, 178187

[43] Guerra, D., Pacheco M. T. and Federizzi L. C., 2013, Phenotypic, cytogenetic and spike fertility characterization of a population of male-sterile triticale. Scientia Agricola, 70(1), 39-47

[44] Bellucci, M., Roscini C. and Mariani A., 2003, Cytomixis in pollen mother cells of *Medicago sativa* L. Journal of Heredity, 94(6), 512-516

[45] Laskar, R. A. and Khan S., 2014, Mutagenic effects of MH and MMS on induction of variability in broad bean (*Vicia faba* L.). Annual Research & Review in Biology, 4(7), 1129